\documentclass[article]{aa} 
\usepackage{adjustbox}
\usepackage{color}
\usepackage{xcolor}
\usepackage{graphicx}
\usepackage{amsmath}
\usepackage{amssymb}
\usepackage{breqn}
\usepackage{gensymb}
\usepackage{placeins}
\usepackage{natbib}
\usepackage{upgreek}
\usepackage{subfig}
\usepackage{multirow}
\usepackage{booktabs}
\usepackage{colortbl}
\usepackage{csvsimple}
\usepackage[colorlinks=true,
linkcolor=blue,
citecolor=magenta]{hyperref}
\usepackage{float}
\usepackage{txfonts}
\usepackage{comment}

\usepackage[switch]{lineno}
\graphicspath{{./}{figures/}}

\newcommand{\be}{\begin{equation}}
\newcommand{\ee}{\end{equation}}

\begin{document}

\title{Imaging ultracompact objects with radiatively inefficient accretion flows}
\titlerunning{Imaging ultracompact objects with radiatively inefficient accretion flows}
\authorrunning{Saurabh et. al.}

\author{Saurabh \inst{1,2}\fnmsep\thanks{sbhkmr1999@gmail.com\\
${}^{\star}$ Member of the International Max Planck Research School (IMPRS) for Astronomy and Astrophysics at the Universities of Bonn and Cologne},~ 
Parth Bambhaniya \inst{2}\fnmsep\thanks{grcollapse@gmail.com} 
and Pankaj S. Joshi\inst{2} \fnmsep\thanks{psjcosmos@gmail.com} 
}

\institute{
        Max-Planck-Institut f\"ur Radioastronomie, Auf dem H\"ugel 69, D-53121 Bonn, Germany
	\and 
        International Centre for Space and Cosmology, School of Arts and Sciences, Ahmedabad University, Ahmedabad-380009, Gujarat, India
        }

\date{}

\abstract

\abstract
{
Recent Event Horizon Telescope observations of M\,87* and Sgr\,A* strongly suggest the presence of a supermassive black hole at their respective cores. We use the semi-analytic radiatively inefficient accretion flows (RIAF) model to investigate the resulting images of the Joshi-Malafarina-Narayan (JMN-1) naked singularity and the Schwarzschild black hole. 
}
{ 
We chose the JMN-1 naked singularity model and compared the synchrotron images with the Schwarzschild solution to search for any distinct features that can distinguish the two objects and to find an alternative to the solution with a black hole.
}
{
We performed general relativistic ray-tracing and radiative transfer simulations using the \texttt{Brahma} code to generate synchrotron-emission images using the thermal distribution function for emissivity and absorptivity. We investigated effects in the images by varying the inclination angle, the disk width, and the frequency. 
}
{
The shadow images simulated with the JMN-1 model closely resemble those generated by the Schwarzschild black hole. The disparities between the two images are very small. We conducted simulations using various plasma parameters, but the resulting images remained largely consistent for both scenarios. This similarity is evident in the horizontal cross-sectional brightness profiles of the two scenarios. Notably, the JMN-1 model exhibits a slightly higher intensity than the Schwarzschild black hole.
}
{ 
We conclude that JMN-1 is a viable substitute for the black hole scenario. This conclusion is not solely grounded in the fact that the two scenarios are indistinguishable from their respective shadow observations, but also in the consideration that JMN-1 emerges as an end state of a continual gravitational collapse. This paradigm not only allows for constraints on spacetime, but also provides a good probe for the nature of the central compact object.
}
\bigskip

\maketitle

\section{Introduction}

In 1939, Oppenheimer, Snyder, and Datt \citep{1939PhRv...56..455O} derived the first dynamical collapse solution for a spherically symmetric and homogeneous dust cloud. This was later known as OSD collapse. Their groundbreaking work demonstrated that a spherically symmetric and homogeneous dust collapse inevitably leads to the formation of a Schwarzschild black hole (BH). This result aligns with Birkhoff's theorem, which states that the Schwarzschild spacetime represents the most general vacuum solution of Einstein's equations. However, the OSD collapse model relies on several idealized assumptions, such as the homogeneous density of the collapsing star, the neglect of gas pressure, and no rotation, to simplify the complex Einstein equations \citep{2011IJMPD..20.2641J}. Although this model unveils the development of an event horizon during the collapse, which prevents the escape of matter, particles, or light to distant observers, it necessitates the exploration of more realistic scenarios that go beyond these idealistic assumptions. By considering the physical complexities and incorporating factors such as nonhomogeneous density profiles and gas pressure, we can expand our understanding of gravitational collapse and its implications.

Large astrophysical bodies typically do not possess a homogeneous density profile throughout their structure. Instead, their density can vary, with a higher density at the core and decreasing values toward the surface. When more physically realistic scenarios are considered that introduce inhomogeneity into the matter distribution profile and account for nonzero pressures, the dynamical collapse can result in the formation of a strong-curvature singularity within the framework of general relativity \citep{2011IJMPD..20.2641J,1993PhRvD..47.5357J,2011CQGra..28w5018J,1968PhRvL..20..878J,2020PhRvD.101d4052M,2021MNRAS.504.4743M}. The final states of these collapses depend on the initial conditions of the collapsing matter.

The general theory of relativity predicts that spacetime singularities necessarily form when sufficiently massive objects collapse under their own gravity. However, the simultaneous formation, or absence, of an event horizon is not enforced. A singularity that forms without an event horizon is known as a naked or cosmic singularity. Naked singularities are intriguing, horizonless compact objects characterized by ultrahigh density regions, where physical quantities diverge arbitrarily. This ultradense compact region can be described by models of various astrophysical compact objects, for instance, a BH, a naked singularity (NS), or a boson star \citep{2011IJMPD..20.2641J,1993PhRvD..47.5357J,2011CQGra..28w5018J,1968PhRvL..20..878J,2003CQGra..20R.301S}. These compact objects have fascinating physical and geometrical properties, and it would be observationally significant to distinguish them.  

The cosmic censorship conjecture (CCC) proposed by Penrose \citep{1965PhRvL..14...57P} posits that strong horizonless singularities are prohibited. However, several research studies have revealed the potential formation of naked singularities during continuous gravitational collapses of inhomogeneous matter clouds. In a seminal work, Joshi, Malafarina, and Narayan (JMN) \citep{2011CQGra..28w5018J} demonstrated that nonzero tangential pressure can prevent the formation of trapped surfaces around the high-density core of a collapsing matter cloud, leading to the emergence of a central NS in extended comoving time. This NS, occurring in physically realistic astrophysical scenarios, holds significant observational consequences. It gives rise to captivating phenomena such as distorted shadows, intense gravitational lensing, and orbital precession of nearby stars \citep{Vagnozzi:2022moj,Sahu:2012er,Bambhaniya:2022xbz}.

On the other hand, black holes have astonishing features as well. As its distinct signature, a black hole possesses an event horizon, which is a one-way membrane in spacetime through which things can fall in, but not escape, and even light cannot escape. Almost every galaxy is thought to have a supermassive black hole at its core \citep{2013ARA&A..51..511K}. There is no conclusive proof at this time for the existence of an event horizon, however. The Event Horizon Telescope (EHT) captured the first shadow image in 2019 in the galalxy M\,87 \citep{2019ApJ...875L...5E}, and more recently, in the center of our own galaxy, Sgr\,A*. \citep{2022ApJ...930L..12E}. While these results are indeed exciting, they should be interpreted with caution. For instance, \cite{2018NatAs...2..585M} and \cite{2023A&A...671A.143R} performed general relativistic magnetohydrodynamics (GRMHD) simulations of a dilated BH and compared it with the Kerr BH. They concluded that it is difficult to distinguish them based on the shadow images alone. Additionally, many other such works indicated the presence of such `mimickers'. Therefore, the shadow images of M\,87* and the Milky Way galactic centers observed by the EHT do not confirm the existence of an event horizon and hence a supermassive black hole (SMBH). Other compact objects can also cast similar shadows. The shadows cast by compact objects such as black holes, naked singularities, grava-stars, and wormholes have been extensively studied in \cite{2019PhRvD.100b4018G,2019PhRvD.100b4020V,Bambhaniya:2021ugr,2019PhRvD.100b4055G,2015PhRvD..92h4005A,  2015PhRvD..91l4020O, Bambhaniya:2021ybs,2019EPJC...79...44S, 2014PhRvD..90j4013S, Solanki:2021mkt,2023ApJ...942...47Y}.

It is generally thought that the shadows arise due to the presence of a photon sphere. Recently, however, \cite{2020PhRvD.102b4022J} have introduced a new spherically symmetric NS solution of the Einstein field equation that lacks a photon sphere, but still casts a shadow. The general criteria for a shadow to occur in the absence of a photon sphere are then determined for null and time-like naked singularities, where both types of singularities satisfy all the energy conditions \cite{2021PhRvD.103b4015D}. Moreover, another model of a NS that was proposed in 1968 by Janis-Newman-Winicour (JNW) can cast a shadow similar to that of a Schwarzschild black hole \cite{2020EPJC...80.1017G,2020PhRvD.102f4027S}. The JNW NS is a minimally coupled massless scalar field solution of the Einstein field equations. \cite{2020PhRvD.102f4027S} showed that as the scalar field charge $q$ increases, the radius of the photon sphere $r_{\mathrm{ph}}$ increases while the shadow radius decreases.

 \cite{2019MNRAS.482...52S} investigated the shadows of JMN-1 NS spacetimes and compared the results to the shadow of a Schwarzschild black hole. For the range of the characteristic parameter $0<M_0<2/3$, there is no photon sphere, and hence no shadow is formed in JMN-1 NS, but an intriguing, full moon image is formed. However, JMN-1 can cast a shadow for $2/3<M_0<1$, similar to that of a Schwarzschild black hole.

The rotating black holes or naked singularities describe the observational data for Sgr A* much better than the static solutions. For rotating spacetimes, the shadow is slightly asymmetric along the spin axis, which will be smaller in diameter than in the Schwarzschild case. The inclusion of spin in the black holes and naked singularities will give the prolate contour shape of the shadow \citep{Bambhaniya:2021ybs}, although for some parameter space that includes spin and scalar field charge, naked singularities will give an arc-shaped shadow \citep{Solanki:2021mkt}. However, for simplicity, we restricted ourselves to static spherically symmetric metrics here, that is, we neglected the effect of spin. The reason for this choice is twofold. First, the effect of spin on the shadow radius is small for the case of a Kerr BH (see, e.g.,  \cite{Vagnozzi:2022moj}).
Second, and perhaps most importantly, there is currently no clear consensus on the spin and inclination angle of Sgr A*. The EHT images are in principle consistent with high spin and low inclination angle, but they are far from being inconsistent with low spin and high inclination angle~\citep{2022ApJ...930L..12E,2022ApJ...930L..16E}. To be precise, however, this is strictly speaking only true for the Kerr metric, which was assumed to derive these results. Independent works based on radio, infrared, and X-ray emission as well as millimeter VLBI, excluded extremal spin ($1-a_{\star} \ll 1$), but were unable to place strong constraints otherwise~\citep{2011ApJ...738...38B, Kato:2009zw, 2009ApJ...697...45B}. Estimates based on semi-analytical models, magnetohydrodynamics simulations, or flare emissions have reported constraints across a wide range of spin values~\citep{Huang:2011qy,2012ApJ...755..133S}. One of the most recent dynamical estimates of the Sgr A$^*$ spin was reported in ~\cite{Fragione:2020khu}, based on the impact on the orbits of the S stars of frame-dragging precession, which would tend to erase the orbital planes in which the S stars formed and are found today: Observations of the alignment of the orbital planes of the S stars today require the spin of Sgr A$^*$ to be very low, $a_{\star} \lesssim 0.1$ (see also ~\cite{Fragione:2022oau}). Some of these estimates rely on an assumed metric, although it is worth pointing out that the result of ~\cite{Fragione:2020khu} does not. Constraints on the inclination angle are even more uncertain. The inconsistency for different estimates of the Sgr A$^*$ spin prompts us to choose a conservative approach in which we neglect the effect of spin, while taking the very recent estimate of ~\cite{Fragione:2020khu} as an indication that the spin may be low: for $a_{\star} \lesssim 0.1$. Additionally, the effect of spin on the shadow radius is negligible at all inclination angles. Moreover, we lack a complete rotating solution of JMN-1 naked singularity so far. Therefore, we restricted ourselves to static spherically symmetric metrics here.

\begin{figure*} 
\centering
\includegraphics[width=\textwidth]{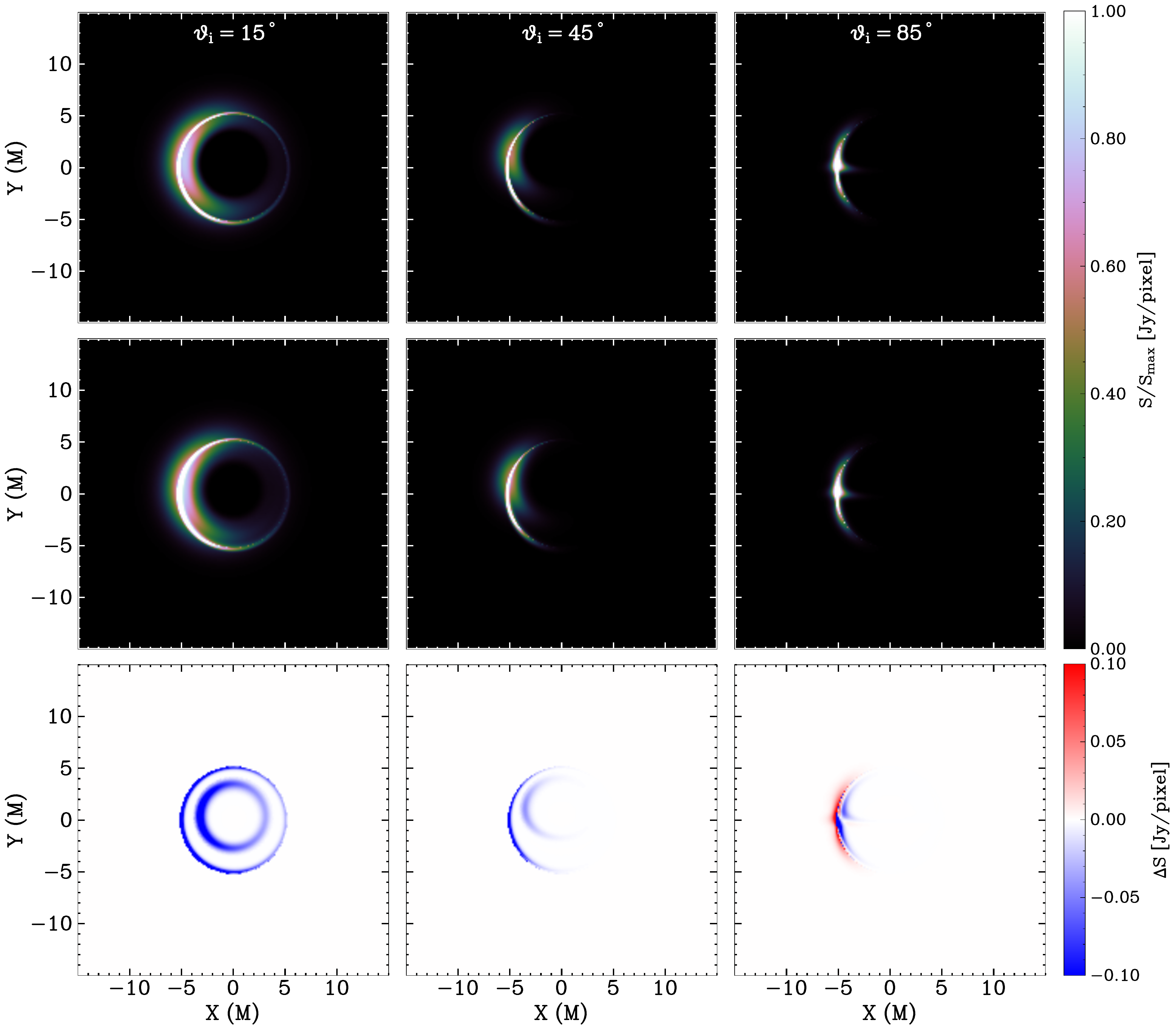}
\caption{Model images for the Schwarzschild BH (first row) and the JMN-1 model (second row) for varied inclination angles ($\theta_{obs}$). The third row corresponds to the various differences in images in the second row from the first row.}
\label{fig:model_inc}
\end{figure*}

\begin{figure} 
\centering
\includegraphics[width=\columnwidth]{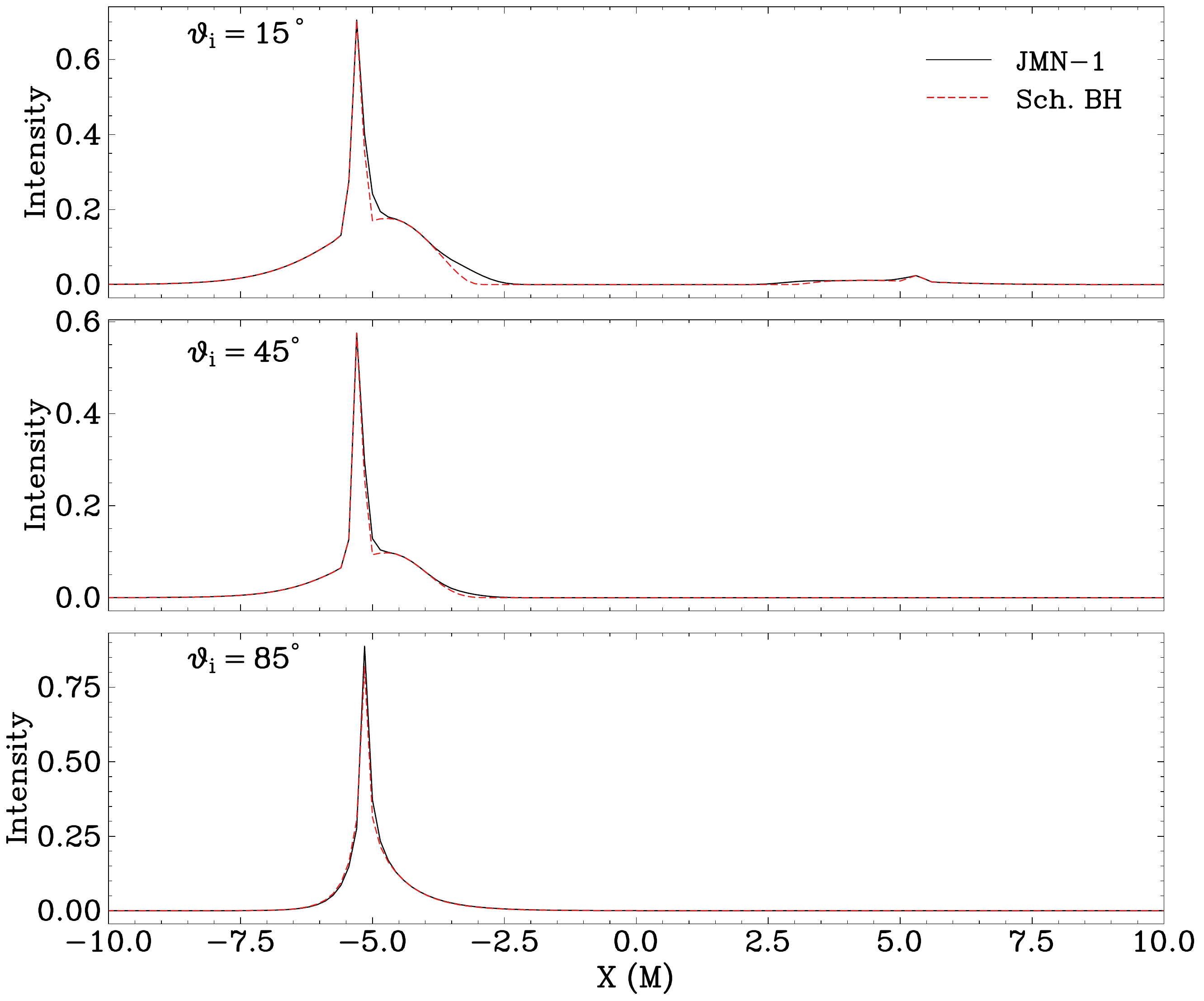}
\caption{Cross-sectional horizontal intensity for the model images of Figure~\ref{fig:model_inc} with varied inclination angles ($\theta_{obs}$).}
\label{fig:cross_inc}
\end{figure}


We used a semi-analytic RIAF model to investigate the resulting images of JMN-1 and the Schwarzschild black hole surrounded by accretion flows. Motivated by RIAF, we integrated the system of geodesics and its radiative transfer equation to compute the theoretically observed intensity.

The paper is organized as follows. In section (\ref{singularity}), we briefly introduce the JMN-1 model. We describe the semi-analytic RIAF model in Section~(\ref{sec:level1}). In Section~(\ref{sec:rtf}), we describe the ray-tracing formalism and the initial conditions we used to generate accretion images. In Section~(\ref{sec:results}), we summarize our results. Finally, the discussion and conclusion are given in Section~(\ref{sec:conclusion}).
\begin{figure*} 
\centering
\includegraphics[width=\textwidth]{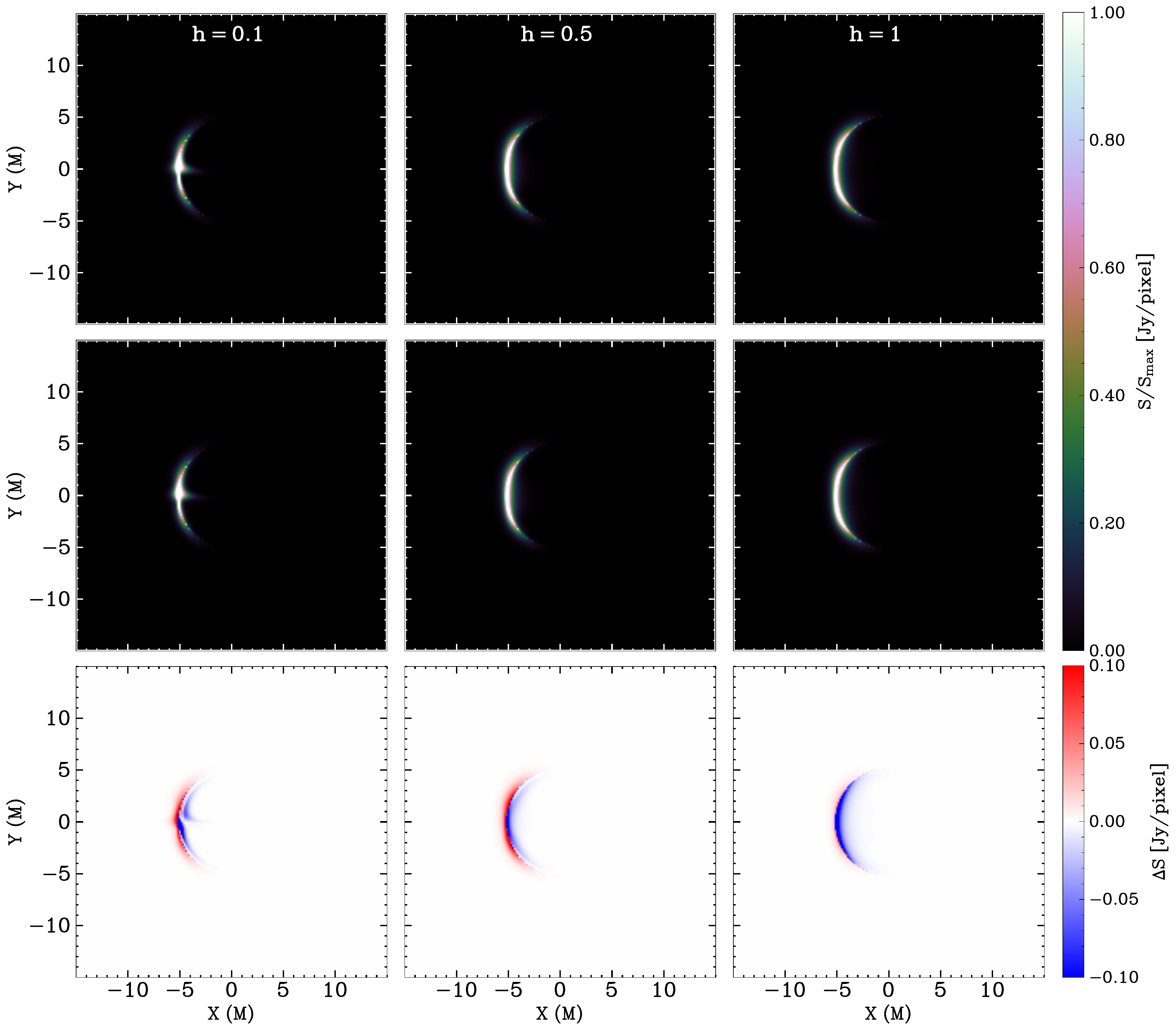}
\caption{Model images for the Schwarzschild BH (first row) and the JMN-1 model (second row) for a varied disk width ($h$). The third row corresponds to the difference in the images in the second row from the first row.}
\label{fig:model_height}
\end{figure*}
\begin{figure} 
\centering
\includegraphics[width=\columnwidth]{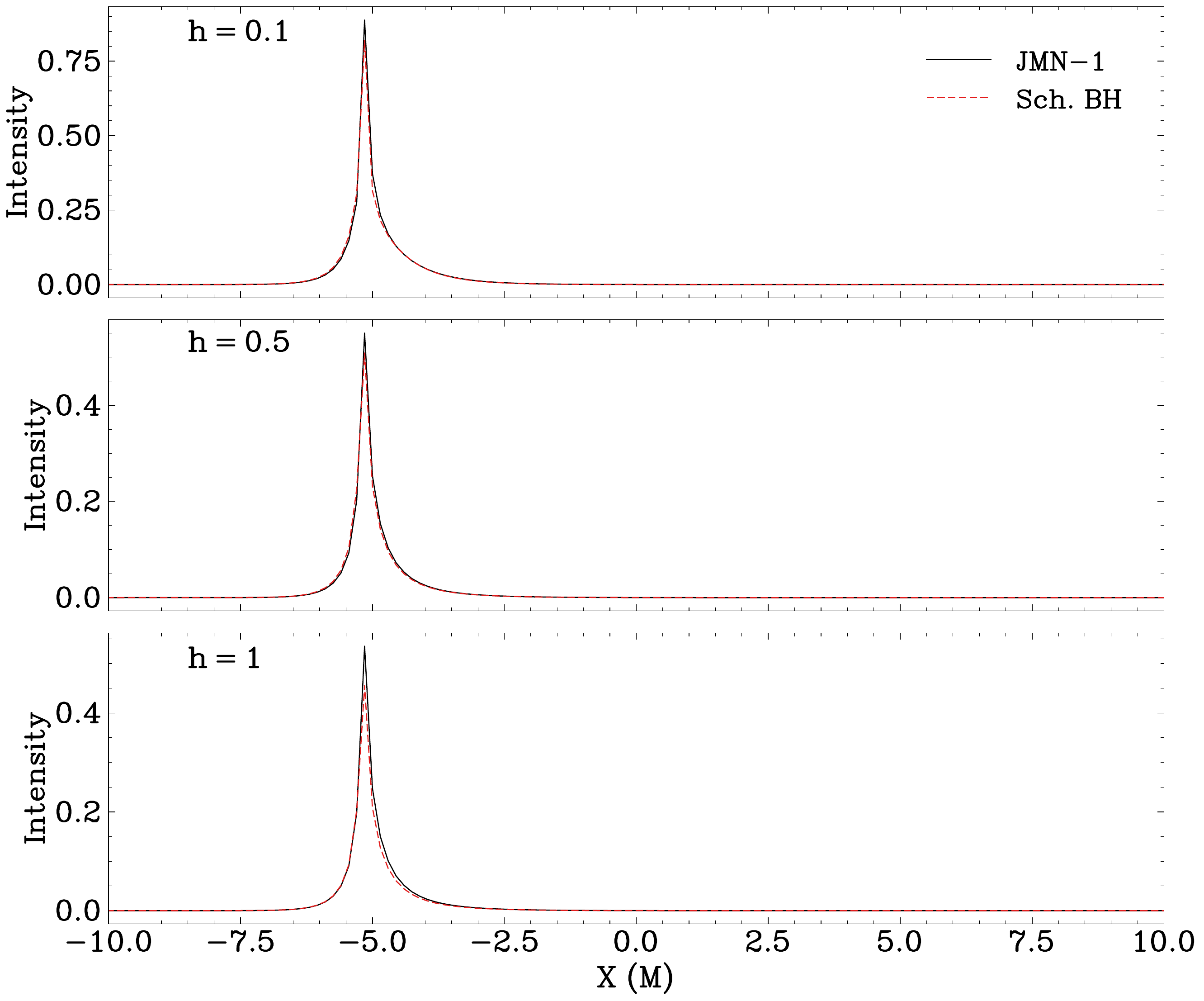}
\caption{Cross-sectional horizontal intensity for the model images of Figure~\ref{fig:model_height} with a varied disk width ($h$).}
\label{fig:cross_height}
\end{figure}

\section{JMN-1 naked singularity}
\label{singularity}
We briefly review the solution that was proposed in the literature for the JMN-1 NS. The spacetime metric for a static and spherically symmetric object in the generalized form can be written as
\begin{eqnarray}
ds^2 = -g_{tt}(r)dt^2 + g_{rr}(r)dr^2 + g_{\theta\theta}(r)d\theta^2 + g_{\phi\phi}(r)d\phi^2,
\end{eqnarray}
where $g_{\mu\nu}$ are the metric tensor components, all of which are only functions of $r$.
The JMN-1 naked singularity can be formed as an end state of gravitational collapse with zero radial pressure and nonzero tangential pressures. These equilibrium configurations of the collapsing fluid within the general theory of relativity were extensively studied in \cite{2011CQGra..28w5018J}.  The JMN-1 naked singularity spacetime is described by the following metric tensor components: 
\begin{eqnarray}
   g_{tt}^{jmn} &=& (1-M_0)\left(\frac{r}{R_b}\right)^{M_0/(1-M_0)},\\
  g_{rr}^{jmn} &=& (1-M_0)^{-1},\\
   g_{\theta\theta}^{jmn} &=& r^2,\\
   g_{\phi\phi}^{jmn} &=& r^2\sin^2{\theta},
   \label{JMN1metric} 
\end{eqnarray}
where $M_0$ and $R_b$ are positive constants. Here, $R_b$ represents the radius of the distributed matter around the central singularity, and $M_0$ should be within the range $0<M_0<1$. The JMN-1 spacetime contains a curvature singularity at $r = 0$. The stress-energy tensor of the JMN-1 spacetime gives the energy density $\rho$ and pressures $p$ as 
\begin{equation}
    \rho=\frac{M_0}{r^2}, \; \; \;
    p_r=0, \; \; \;
    p_{\theta}=\frac{M_0}{4(1-M_0)}\rho.
\end{equation}
The JMN-1 naked singularity is a nonvacuum solution of the Einstein field equations that satisfies all energy conditions. An equilibrium configuration of a massive anisotropic matter fluid can be achieved as an end state of dynamical gravitational collapse with nonzero tangential pressure. The matter source can be considered as baryonic and/or dark matter.

The spacetime metric is modeled by considering a high-density compact region in a vacuum, which means that the spacetime configuration should be asymptotically flat. Therefore, if any spacetime metric is not asymptotically flat, we must match that interior spacetime to an asymptotically flat exterior spacetime with a particular radius. The latter is the case for the JMN-1 naked singularity. We can smoothly match this interior spacetime to exterior Schwarzschild spacetime at $r=R_b$ as
\begin{equation}
    ds^2 = -\left(1-\frac{M_0 R_b}{r}\right) dt^2+\left(1-\frac{M_0 R_b}{r}\right)^{-1}dr^2+r^2d\Omega^2,
    \label{matchSCH}
\end{equation}
where $\text{d}\Omega^2 = d\theta^2+\sin^2\theta d\phi^2$ and $M=\frac{1}{2}M_0 R_b$ is the total mass of the compact object. The extrinsic curvatures of JMN-1 and Schwarzschild spacetimes are automatically smoothly matched at $r=R_b$ because the JMN-1 spacetime has zero radial pressure \citep{2019PhRvD.100l4020B}. 
The complete solution for the JMN-1 metric in the interior $r < R_b$ is given by
\begin{eqnarray}
ds^2 &=& - (1-M_0)\left(\frac{r}{R_b}\right)^{\frac{M_0}{(1-M_0)}} dt^2 + \frac{dr^2}{(1-M_0)} \\ &+& r^2 d\theta^2 + r^2\sin^2{\theta} d\phi^2 \notag.
\end{eqnarray}
In order to specify the nature of the central singularity, we note that the outgoing radial null geodesics in the spacetime above are given by
\begin{equation}
    \frac{dr}{dt} = (1-M_0)\left(\frac{r}{R_b}\right)^{M_0/2(1-M_0)}.
\end{equation}
It is then easy to confirm that light rays escape from the singularity (for all values of $M_0 < 2/3$). From the
above equation, which gives
\begin{equation}
t(r)=\frac{2 R_b^{\frac{M_0}{2(1-M_0)}}}{2-3M_0}  r^{\frac{2-3M_0}{2(1-M_0)}},
\end{equation}
we immediately see that the comoving time required by a photon to reach the boundary is $t_b = \frac{2R_b}{(2 - 3M_0)} < +\infty $. It
follows that there are future directed null geodesics in the spacetime that reach the boundary of the cloud, and which in the
past terminated at the singularity, thus showing this to be a naked singularity.

\section{Radiative transfer} \label{sec:level1}

In this section, we describe the prescription for the semi-analytic radiative inefficient accretion flows (RIAF) model. We closely follow \cite{2018ApJ...863..148P} for some basic definitions of the parameters.
The covariant form of general relativistic radiative transfer is expressed as  
\begin{equation}
\frac{d\mathcal{I}}{d\tau_\nu} = -\mathcal{I} + \frac{{\eta}}{\chi},
\end{equation}
where $\mathcal{I}$ is the Lorentz-invariant intensity and is related to the specific intensity via $\mathcal{I} = I_\nu/\nu^3 = I_{\nu_0}/\nu_0^3$, where the subscript 0 denotes quantities in the local rest frame. $\tau_\nu$ is defined as the optical depth. $\chi$ and $\eta$ are the invariant absorption coefficient and emission coefficient at frequency $\nu$. These coefficients are defined by the specific emissivity and absorptivity. For this work, we used the thermal electron distribution function (eDF) \citep{2016ApJ...822...34P}. 

The number density and temperature of electrons can be written as a hybrid combination of radial and exponential functions given by
\begin{eqnarray}
    n_{\rm th} &=& n_{\rm e,th}r^{-\alpha}e^{-z^2/2h^2}, \\
    T_{\rm th} &=& T_{\rm e,th}r^{-\beta}, 
\end{eqnarray}
where $z=r\cos\theta$, and $h$ is the disk width. $n_{\rm e,th}$ and $T_{\rm e,th}$ are the respective normalization parameters.  The magnetic field strength of the toroidal field is then given by
\begin{eqnarray}
    \frac{B ^ 2}{8\pi} = \frac{1}{10}n_{\rm th}\frac{m_p c^2}{6r} [r_g],
\end{eqnarray}
where $r_g=GM/c^2$, and $m_p$ is the mass of proton.

We modeled the four-velocity of the plasma accreted onto the central compact object by following \cite{2023ApJ...942...47Y}. The four-velocity can be written as
\begin{equation}
    u^{\mu} = (u^t, u^r, 0, \Omega u^t),
\end{equation}
where  $(\Omega)$ is the angular velocity. The radial component of the four-velocity can be written in parameterized form,
\begin{equation}
    u^{r} = -\eta_r\left(\frac{r}{R_{\rm ISCO}}\right)^{-n_r},
\end{equation}
where $\eta_r = 0.1$ and $n_r=1.5$ are the free parameters. We kept their values fixed for the current analyses.

Because no stable circular orbits are possible beyond the innermost stable circular orbit (ISCO) radius ($r_{\rm ISCO}$), the particles will plunge with the constant energy ($E_{\rm ISCO}$) and the constant angular momentum ($L_{\rm ISCO}$) acquired at the ISCO radius. In this way, we can model the motion of particles for $r<r_{\rm ISCO}$.

\begin{figure*} 
\centering
\includegraphics[width=\textwidth]{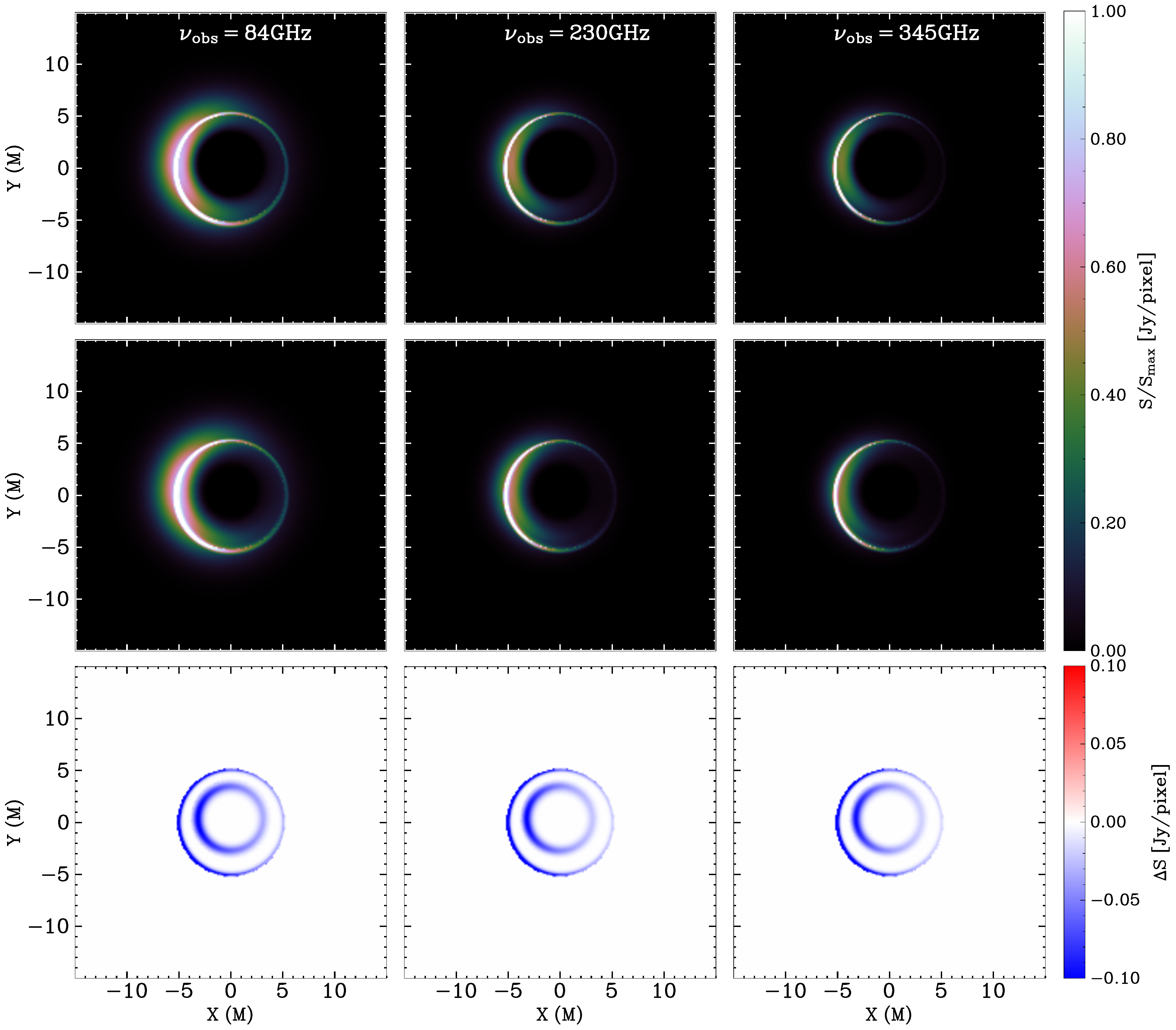}
\caption{Model images for the Schwarzschild BH (first row) and the JMN-1 model (second row) for varied observed frequencies ($\nu_{\rm obs}$). The third row corresponds to the difference in the images in the second row from the first row.}
\label{fig:model_freq}
\end{figure*}

\begin{figure} 
\centering
\includegraphics[width=\columnwidth]{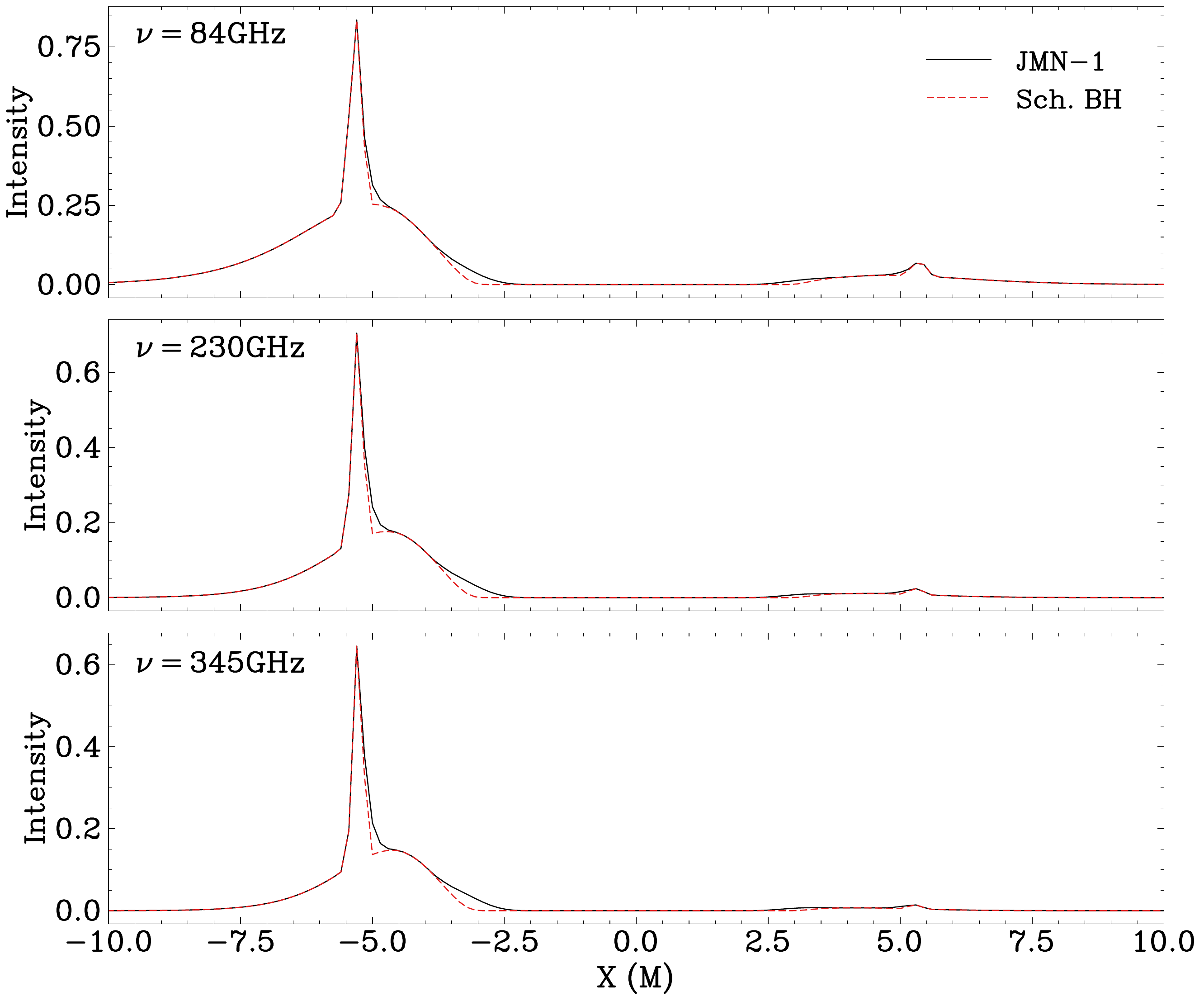}
\caption{Cross-sectional horizontal intensity for the model images of Figure~\ref{fig:model_freq} with varied observed frequencies ($\nu_{\rm obs}$).}
\label{fig:cross_freq }
\end{figure}

\section{Ray-tracing formalism}
\label{sec:rtf}

In order to calculate the shadow image of a BH or a naked singularity, the geodesic equations in the considered background spacetime must first be solved. To this end, we used the \texttt{Brahma} code (\textit{in preparation}). We briefly describe the approach for ray-traced images of the models described in Section~\ref{sec:level1}. We solve a system of six differential equations ($\dot{t}, \dot{r}, \dot{\theta}, \dot{\phi}, \dot{p_r}, \dot{p_\theta})$.

\subsection{Geodesic equations of motion}

For a given metric $g_{\alpha \beta}$, the Lagrangian and the Hamiltonian can be written as
\begin{eqnarray}
2\mathcal{L} &=& g_{\alpha \beta}\,\dot{x}^{\alpha}\dot{x}^{\beta}, \\
2\mathcal{H} &=& g^{\alpha \beta}p_{\alpha}p_{\beta},
\end{eqnarray}
where an overdot denotes differentiation with respect to the affine
parameter, $\lambda$. The corresponding equations of motion are given by 
\begin{eqnarray}
    \dot{x}^\alpha &= \frac{\partial H}{\partial p_{\alpha}}, \\
    \dot{p}_\alpha &= -\frac{\partial H}{\partial x^{\alpha}}.
\end{eqnarray}
From the Lagrangian, the covariant four-momenta of a geodesic may be written as
\begin{eqnarray}
p_\alpha = \frac{\partial \mathcal{L}}{\partial \dot{x}^\alpha}.
\end{eqnarray}

Using the conservation of energy and angular momentum, this can be broken down to write $p_t = -E$ and $p_\phi=L$, where $E$ is the total energy of the particle, and $L$ is the angular momentum in the direction of $\phi$. 
To solve the system of an ordinary differential equation (ODE), initial conditions are required. Different authors have assumed different initial conditions based on the type of work and ODEs that were to be solved. In \texttt{Brahma}, we used the formalism described in \cite{2016PhRvD..94h4025Y}. The observer is placed far away from the compact object ($r_{\mathrm{obs}} = 10^{3}\,M$), where the spacetime is assumed to be  flat (asymptotic flatness).

\section{Results} \label{sec:results}

We used the formalism described in Section~\ref{sec:rtf} to trace the photons along geodesics and solved for the observed intensity. In principle, the model consists of various parameters that can be altered to study specific properties of the accretion flows. In our simulations, we currently restricted ourselves to a specific set of parameters $\Theta_p = (\theta_{obs}, \nu_{obs}, h)$. We studied the impact of these parameters in the shadow images of the Schwarzschild BH and the JMN-1 naked singularity. The default values for the parameters were $\mathbf{M_0=0.7}$, $\nu_{\rm obs} = 230$\,GHz,\,$h=0.1$,\,$n_{e,th}\approx10^5$ cm$^{-3}$\,and $T_{\rm e,th}\approx10^{10}$\,K.

\begin{itemize}
    \item \textit{Inclination angle}:  The first row in Figure~\ref{fig:model_inc} corresponds to the Schwarzschild BH, the second row corresponds to the JMN-1, and the third row represents the image difference by subtracting second-row images from the first-row images. The first, second, and third columns represent $\theta_{obs} = 15^\circ$, $45^\circ$, and $85^\circ$, respectively. Although the Schwarzschild BH and JMN-1 model images may look identical in terms of shadow features and emission, slight differences are visible in the third row. Figure~\ref{fig:cross_inc} represents the intensity along the horizontal cross section of the images, and the overall profile remains similar throughout the image except for minor changes.
    \item \textit{Disk width}: Figure~\ref{fig:model_height} is the same as Figure~\ref{fig:model_inc}, except for the change in the disk width $h$. The first, second, and third columns correspond to $h=0.1$, $0.5$, and $1$, respectively. Higher and lower values of $h$ allow for a thicker and thinner flow geometry of the accreting matter, which would correspond to a transformation from a spherical-like flow into a disk-like flow. The inclination angle was set to $85^\circ$.  For $h=0.1$, a distinct disk-like geometry is visible, but this does not hold for the other cases. Because we used a thermal eDF, the images are different from those presented in \cite{2018ApJ...863..148P}. For different values of $h$, the images are very similar for the Schwarzschild BH and the JMN-1 model, and the corresponding horizontal cross section is shown in Figure~\ref{fig:cross_height}. The profiles for both spacetimes are again similar here, with JMN-1 having a slightly higher intensity for $h=0.1$.
    \item \textit{fFequency}: Currently, the EHT operates at $230$ GHz, and there are plans for including other frequency bands as part of EHT and ngEHT. We therefore also performed simulations for different frequencies, and the results are plotted in Figure~\ref{fig:model_freq} for $84$ GHz, $230$GHz, and $345$ GHz. No significant changes are observed for the two spacetimes relative to each other when the observed frequency is changed. As the frequency is increased, however, the emission becomes thinner, which agrees with previous results. The horizontal cross-sectional intensity as plotted in Figure~\ref{fig:cross_freq } retains a similar profile as in previous cases for different values of the disk width and inclination angle.
\end{itemize}

\section{Conclusion and discussion}\label{sec:conclusion}

We have modeled radiative inefficient accretion flows around the JMN-1 cosmic singularity and calculated the resulting shadow images. Because JMN-1 causes a shadow of the same size as the Schwarzschild BH, it is worthwhile to investigate various effects from the change in spacetime on the accretion flows and the resulting images. We first defined the geodesic equations for both photons and particles. We then defined our semi-analytic RIAF model in Section~\ref{sec:level1}. Thereafter, we solved the system geodesic equations with the ray-tracing formalism as described in Section~\ref{sec:rtf} simulatenously with the radiative transfer equation to calculate the observed intensity. In order to compare the JMN-1 naked singularity with the Schwarzschild BH, we carried out simulations to generate images for various values of the parameters: for the inclination angle ($\theta_{obs}$), the disk width ($h$), and the observed frequency ($\nu_{obs}$). The results from the simulations are presented in Section~\ref{sec:results}.

The visual representations of the Schwarzschild BH and JMN-1 model exhibit striking similarities. Subtle differences lie in the slightly heightened intensities observed for the JMN-1 model. Even though we varied several parameters associated with the accretion disk, the overall characteristics of the disk remained relatively unchanged. This reflects the results of previous studies \citep{2019MNRAS.482...52S, 2019JCAP...10..064S}. The results we obtained here contribute to an expanded understanding of a realistic model for the accretion disk, which is astrophysically relevant in particular in the context of AGNs. Following the fact that JMN-1 emerges as the end state of gravitational collapse and acts as galactic spacetime, this opens up an intriguing avenue in the study of AGNs in which the central engine powering the galaxy is a visible cosmic singularity. This analysis concludes that if horizon-scale observations of the central supermassive compact object produce a shadow-like structure, it is impossible to distinguish between a BH and a naked singularity, and by extension, this is valid for the recent studies of M\,87* and Sgr\,A*. 

However, many other paths can be pursued further in order to test and investigate the robustness of the model. For example, we can generate synthetic VLBI observations from the ray-traced images presented in this paper to directly compare our results to observational data.

\begin{itemize}

    \item For instance, \cite{2019MNRAS.482...52S} have considered a spherically accreting system and studied \textbf{the corresponding} shadow images. Bondi accretion flow was also used to study the shadow and spectral properties in MHD simulations. This treatment can also be applied to our model to study the spectra, which can then be used to study specific sources such as Sgr A*. The spectral signatures will be studied in an accompanying paper. 

    \item We only studied the horizon-scale accretion-flow images of the models, and in GRMHD simulations, black holes produce highly energetic jets that were variously observed. Studies of jets and accretion flows in GRMHD simulations \citep{2023arXiv230715140K} will help us to conclude whether the Universe contains horizonless objects.

\end{itemize}

Using the models described in this work, it will be possible to unveil the true nature of the objects in M\,87* and Sgr\,A*. From the studies so far, it would appear that naked singularities present a possible alternative to the standard picture of a black hole within the framework of nonmodified general relativity.

\begin{acknowledgements}
Saurabh received financial support for this research from the International Max Planck Research School (IMPRS) for Astronomy and Astrophysics at the Universities of Bonn and Cologne. We would also like to acknowledge Jan R\"oder for his role as the internal referee at the MPIfR and helpful comments.
\end{acknowledgements}
\bibliography{ref}{}
\bibliographystyle{aa}

\end{document}